\documentclass[prl,twocolumn]{revtex4}

\usepackage{graphicx}% Include figure files
\usepackage{dcolumn}% Align table columns on decimal point
\usepackage{bm}% bold math

\begin{document}

\title{Surface segregation in nanoparticles from first principles}

\author{Roman V. Chepulskii$^{1}$, W. H. Butler$^{2}$, A. van de Walle$^3$, and Stefano Curtarolo$^{1}$}

\affiliation{$^{1}$Department of Mechanical Engineering and Materials Science, Duke University, Durham, NC 27708\\
$^{2}$Center for Materials for Information Technology, University of
Alabama, Tuscaloosa, AL 35487\\
$^{3}$ Division of Engineering and Applied Science, California
Institute of Technology, Pasadena, CA 91125}

\begin{abstract}

FePt nanoparticles are known to exhibit reduced L1$_0$ order with
decreasing particle size. The reduction in order reduces the magnetic anisotropy and the thermal stability of the direction of magnetization of the particle.
The phenomenon is addressed by investigating the thermodynamic
driving forces for surface segregation using a local (inhomogeneous)
cluster expansion fitted to \emph{ab initio} data which accurately
represents interatomic interactions in both the bulk and surface
regions. Subsequent Monte Carlo simulations reveal that first
surface layer Pt segregation is compensated by Pt depletion in the
second subsurface layer. This indicates that the core's ordered
state is not affected by surface thermodynamics as much as
previously thought. Thus, the weak ordering experimentally observed
is likely not due to fundamental thermodynamic limitations but
rather to kinetic effects.

\end{abstract}


\maketitle

Understanding the physics of the transition from bulk to nanoscale
magnetic alloys is important from both a fundamental \cite{Fundam}
and technological point of view. In the last decade, Fe-Pt
nanoparticles have been intensively investigated in view of possible
future applications as an ultra-high density information-storage
medium and high-performance permanent magnets
\cite{FePt-nano,NanoSize,Liu-Harrell}. The critical issue for
information-storage application is the presence of magnetic
anisotropy offering sufficiently large thermal stability. In Fe-Pt,
a high magnetic anisotropy is guaranteed by an ordered L1$_0$ phase.
However, recent experimental observations have shown a difficulty in
obtaining a high degree of L1$_0$ order in FePt nanoparticles annealed
at $T\lesssim$600$^{\circ}$C with diameter less than $\sim$ 4nm
\cite{NanoSize,Sint}. Due to the high surface to volume ratio of
nanoparticles, surface segregation has been suggested to be one of
the possible causes of this reduced ordering
\cite{RCWB-05,Asta05-06,Albe05-07} (interestingly, in FeC and FeMoC
nanoparticles, size plays the opposite role inducing a disorder-order
transition \cite{art36-39}).

In this letter, we address the FePt surface segregation problem with
accurate \emph{ab initio} computational thermodynamics techniques.
In earlier studies, the phenomenon was tackled using several
interatomic model potentials \cite{Asta05-06,Albe05-07,Wiest07}.
Pair interactions were fitted to the experimental phase diagram \cite{Asta05-06,Albe05-07},
and to bulk first principles data \cite{Asta05-06}.  A single-layer isotropic surface potential was
estimated from the surface energy difference between pure fcc Fe and
Pt \cite{Asta05-06} and by fitting to experimental segregation
profiles \cite{Albe05-07}. The Embedded Atom Method potential was
used in Ref. [\onlinecite{Wiest07}]. More recently, surface segregation in small
Fe-Pt clusters was studied by direct comparison of the energies of
several different configurations obtained from first principles
\cite{Gruner08}. Our approach is radically different: the surface
potential is obtained from \emph{ab initio} calculations without
fitting to experimental data and without \emph{a-priori} assumptions
about strength and isotropy. We combine the surface potential
obtained using the method just described and the bulk pair potential \cite{RCWB-05} into a local
(inhomogeneous) cluster expansion \cite{LCEs}, enabling efficient
Monte Carlo simulations to describe the nanoparticle's actual
thermodynamic equilibrium segregation profile. The temperature and
size dependencies of the L1$_0$ order identify the regimes when
surface segregation is responsible for reduced equilibrium order
with correspondingly low magnetic anisotropy.

\begin{figure}[ht]
%  \vspace{-3mm}
  \includegraphics[width=0.40\textwidth]{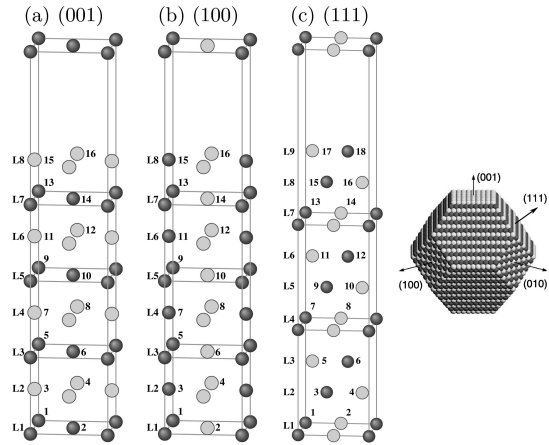}
%  \vspace{-7mm}
  \caption{The unit cells of crystal structures representing atomic
    slabs separated by vacuum used to model the nanoparticle's facets.
    Each slab is obtained from an unrelaxed L$1_0$ crystal structure along (a) (001), (b) (100), and (c) (111) planes.
    Layers (L) and atoms are labeled with numbers. Fe and Pt atoms are represented as black and grey circles, respectively.} \label{FIG:slabs}
\end{figure}

\begin{figure}[ht]
  \includegraphics[width=0.38\textwidth]{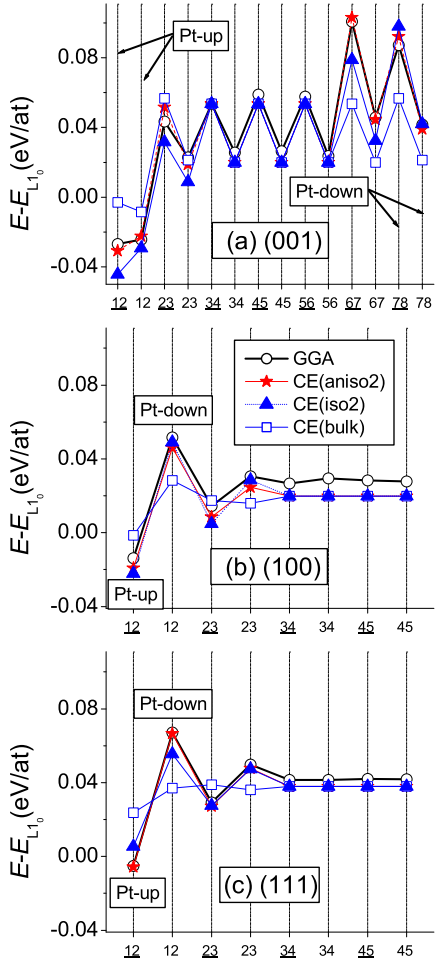}
%  \vspace{-7mm}
  \caption{(Color online) Formation energies of different slabs with respect to ideal L$1_0$ slabs from both first principles
    (GGA) and using three cluster expansions ``aniso2'', ``iso2'', and
    ``bulk'' (see Tab. \ref{TAB:CEs}). The slabs are identified by the label ``$ij$'', where $i$ and $j$ indicate the layers
    between which the atoms are exchanged (Fig. \ref{FIG:slabs}).
    Two configurations are considered: $ij$ and $\underline{ij}$ for every choice of $i$ and $j$.
    ``Pt-up''/''Pt-down'' mean that Pt atoms move
    to/from the surface of the corresponding perfect L$1_0$ slab, respectively.} \label{FIG:E_i}
\end{figure}

\begin{figure}[ht]
  \includegraphics[width=0.38\textwidth]{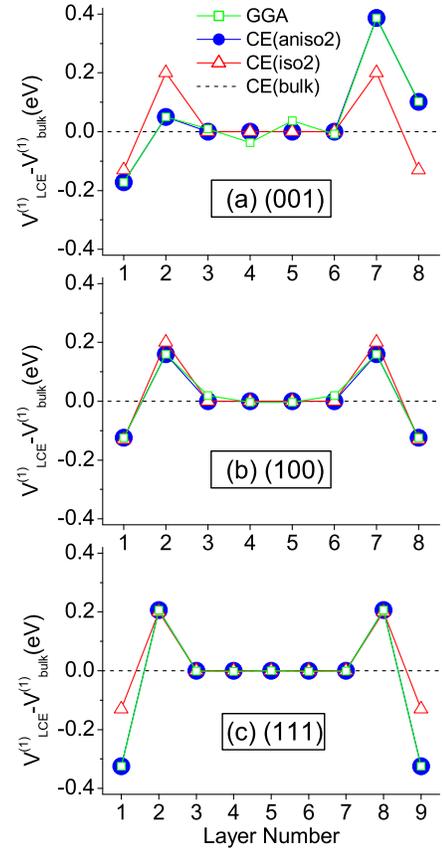}
%  \vspace{-8mm}
  \caption{(Color online) The dependence of surface potential
    $V^{(1)}_\texttt{LCE}-V^{(1)}_\texttt{bulk}$ on the layer number in
    (a) (001), (b) (100), and (c) (111) slabs corresponding to initial
    first principles (GGA) data and three cluster expansions: "aniso2",
    "iso2", and "bulk" (see Tab. \ref{TAB:CEs}).} \label{FIG:Mu_i}
\end{figure}

The nanoparticle is modeled as an fcc-based ``truncated octahedron''
(Fig. \ref{FIG:slabs}), which is typically observed for chemically
synthesized FePt nanoparticles \cite{Fundam,TO-exper} with free
boundary conditions. The initial atomic configuration is considered
as ordered L1$_0$  with alternating Fe and Pt layers forming (001)
fcc crystal planes. This particle has three inequivalent types of
surfaces, (001), (100), and (111), which are addressed by using
standard periodic slab geometries (Fig. \ref{FIG:slabs}). The vacuum layer thickness
was chosen to be approximately equal to three fcc lattice
parameters. As well as perfectly ordered slabs, defected surfaces
obtained by exchange(s) of atoms belonging to different atomic
layers are considered. The energies were calculated from first
principles within the Generalized Gradient Approximation using Projector Augmented Wave pseudopotentials, as
implemented in the VASP package \cite{VASP,noteVASP}. The calculated
slab energy differences are presented in Fig. \ref{FIG:E_i}. In
agreement with previous theoretical
\cite{Asta05-06,Wiest07,Gruner08,Ruban99-09} and experimental
reports \cite{ExpSegr}, the data indicates that there is an
energetic gain for Pt atoms to segregate into all three considered
surfaces. However, this finding is not sufficiently compelling to
determine the thermodynamic equilibrium of the system.

\begin{table}[ht]
  \caption{Cluster expansions obtained for a description of
    nanoparticle configurational energies: ``bulk'' is the bulk CE
    obtained in \cite{RCWB-05} without accounting for surface effects;
    ``iso1'' and ``iso2'' (``aniso1'' and ``aniso2'') designate the
    implementation of isotropic (anisotropic) surface potential
    affecting one and two external layers, respectively. ``100iso1'' and
    ``100iso2'' are the hybrid CEs for which the surface potential is the
    same for all (001) and (100) types of surfaces but is different from
    the surface potential for the (111) surface. The least-square-fitting (LSF)
    errors are those of atom exchange energies in two external layers
    (total and within each surface).} \label{TAB:CEs} {\small
    \begin{tabular}{cc|ccccccc}
      \hline\hline
      \multicolumn{2}{c|}{Layers} & \multicolumn{7}{c}{Cluster Expansions}  \\
      (lmn)    &   $i$ &   bulk    &   iso1   &   iso2   &   100iso1   &   100iso2   &   aniso1 &   aniso2 \\
      %    &   $i$ &   Bulk    &   1/iso   &   2/iso   &   1/100iso   &   2/100iso   &   1/aniso &   2/aniso \\
      \hline
      & & \multicolumn{7}{c}{$V^{(1)}_\texttt{LCE}-V^{(1)}_\texttt{bulk}$(eV)}  \\
      (001)   &   8   &   0   &   -0.33   &   -0.13   &   -0.264  &   -0.065  &   -0.284  &   0.102   \\
      &   7   &   0   &   0   &   0.201   &   0   &   0.199   &   0   &   0.386   \\
      &   2   &   0   &   0   &   0.201   &   0   &   0.199   &   0   &   0.050   \\
      &   1   &   0   &   -0.33   &   -0.13   &   -0.264  &   -0.065  &   -0.222  &   -0.172  \\
      (100)   &   2   &   0   &   0   &   0.201   &   0   &   0.199   &   0   &   0.160   \\
      &   1   &   0   &   -0.33   &   -0.13   &   -0.264  &   -0.065  &   -0.284  &   -0.124  \\
      (111)   &   2   &   0   &   0   &   0.201   &   0   &   0.207   &   0   &   0.207   \\
      &   1   &   0   &   -0.33   &   -0.13   &   -0.531  &   -0.324  &   -0.531  &   -0.324  \\
      &       &   \multicolumn{7}{c}{LSF Error(eV/at)}                                                   \\
      (001)   &       &   0.026   &   0.021   &   0.013   &   0.020   &   0.012   &   0.020   &   0.004   \\
      (100)   &       &   0.015   &   0.009   &   0.007   &   0.002   &   0.006   &   0.008   &   0.002   \\
      (111)   &       &   0.022   &   0.011   &   0.008   &   0.008   &   0.003   &   0.008   &   0.003   \\
      Total   &       &   0.023   &   0.016   &   0.011   &   0.015   &   0.009   &   0.015   &   0.003   \\
      \hline\hline
  \end{tabular}}
\end{table}

To parameterize the entropic contributions to surface segregation, a
number of local cluster expansions (LCE) are constructed so that they
reproduce atom exchange energies obtained from total energy
calculations both in the bulk and in the surfaces. The local (or
inhomogeneous) nature of the cluster expansions (CE) manifests
itself by the presence of a layer-dependent unary mixing potential
$V^{(1)}$ (see Eqs. (4,7) in \cite{RCWB-05}). Such a layer
dependence near the surface may be formally considered as an
external surface potential applied to the surface atoms. The
constructed LCEs differ from each other by the number of
external layers affected by surface potential and by the directional
dependence of surface potential. The constructed LCEs and their
accuracy in surface regions are presented in Fig. \ref{FIG:Mu_i} and
Tab. \ref{TAB:CEs}. The previously proposed bulk CE \cite{RCWB-05}
has to be modified only at two external surface layers (see Fig.
\ref{FIG:Mu_i}) so that only the LCEs with one and two layers of
surface potentials have to be considered. Comparing the prediction
errors of different CEs, it is concluded that accounting for surface the
potential within the second layer has a larger effect than considering
the anisotropy in the surface potential. Hence, we conclude that the
``iso2'' LCE (with isotropic two-layer surface potential) represents
the best compromise between the accuracy and complexity of
calculations. Note that the ``iso1'' surface potential (-0.33 eV) is
very similar to the corresponding surface potential (-0.30 eV)
obtained in Refs. [\onlinecite{Asta05-06,Albe05-07}] with different
approaches.

\begin{figure}[ht]
  \includegraphics[width=0.5\textwidth]{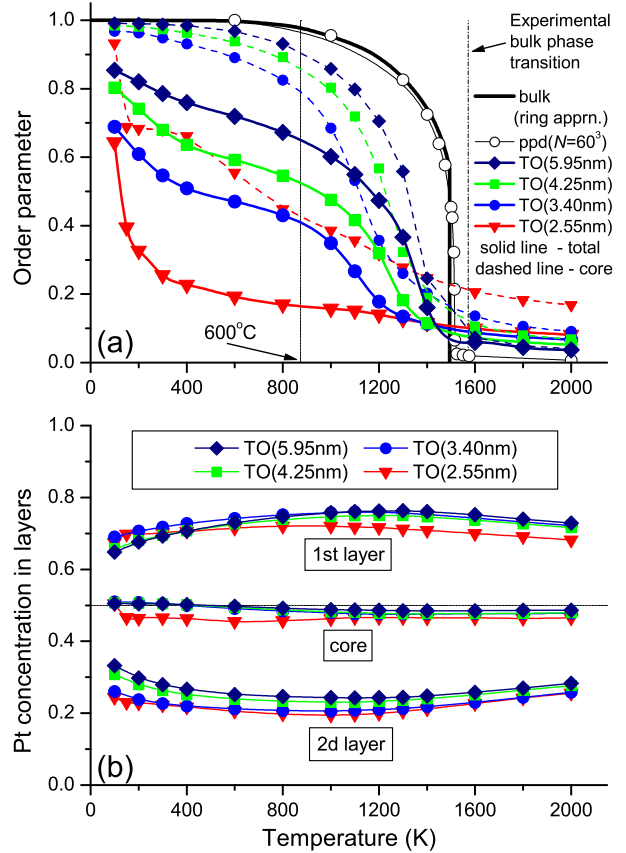}
%  \vspace{-10mm}
  \caption{(Color online) The temperature dependence of the equilibrium
    (a) L1$_0$ order parameter $\overline{\eta}$ and (b) Pt
    concentrations in two external layers and internal core for
    truncated octahedron (``TO'') shape nanoparticles with different sizes
    at near equiatomic composition. In graph (a), we include the
    data obtained within the analytical ring approximation \cite{Ring}
    for bulk (``bulk'') as well as by Monte Carlo simulation for the
    parallelepiped (``ppd'') sample containing $N=60^3$ atoms. The solid
    and dashed lines represent the total and core (excluding two
    external layers) order parameters, respectively.} \label{FIG:PP_T}
\end{figure}

\begin{figure}[ht]
  \includegraphics[width=0.5\textwidth]{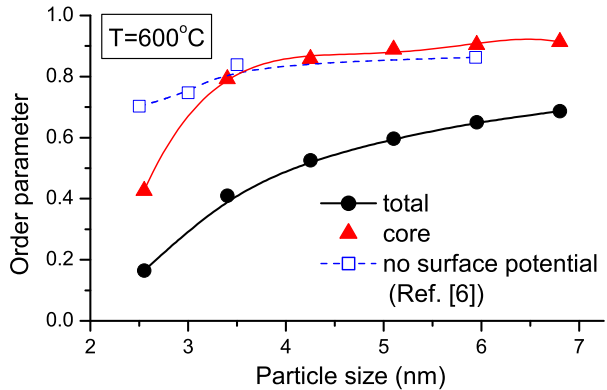}
 % \vspace{-10mm}
  \caption{(Color online). Equilibrium L1$_0$ order
    parameters $\overline{\eta}$ versus particle size at
    $T=$600$^{\circ}$C: total, core (excluding two external layers) and
    obtained in Ref. \cite{RCWB-05} by neglecting surface potential. }
  \label{FIG:PP_d}
\end{figure}

The cluster expansion is then employed in a finite-temperature Monte Carlo
simulation scheme \cite{RCWB-05}. While spanning the phase space,
the L1$_0$ order is monitored by the  ``generalized'' parameter $\overline{\eta}$:
\begin{equation} \label{EQ:PP}
  \overline{\eta}=\left\langle \max \left\{ |\eta_x|,|\eta_y|,|\eta_z|
  \right\}-\min \left\{ |\eta_x|,|\eta_y|,|\eta_z| \right\}
  \right\rangle,
\end{equation}
where $\left\langle\ldots\right\rangle$ is the statistical average
over the Monte Carlo steps, and three directional parameters
$\eta_i (i=x,y,z)$ are defined as the difference between the Pt atom
concentrations at odd and even crystal planes perpendicular to the
\emph{i}-th direction.
The generalization is required for small nanoparticles, because, in
the presence of a surface potential, we have observed that all $\eta_i$
can be comparable in wide temperature intervals.
Generally, we can not neglect $\min
\left\{ |\eta_x|,|\eta_y|,|\eta_z| \right\}$ in Eq. (\ref{EQ:PP}) as
it was done Ref. [\onlinecite{RCWB-05}].
In the case of isotropic order, $|\eta_x|=|\eta_y|=|\eta_z|$,
we have $\overline{\eta}=0$.

The results of our Monte Carlo simulations with the ``iso2'' cluster
expansion are presented in Figs. \ref{FIG:PP_T}-\ref{FIG:PP_d}.
The comparison of the present and previous \cite{RCWB-05} results
confirms that the presence of a surface potential
reduces the {\it total} order in agreement with Refs. [\onlinecite{Asta05-06,Albe05-07}].
The reduction is larger for
smaller particles having a bigger fraction of surface versus volume.
In addition, there is a negligible effect of the surface potential on the
core order parameter for particles larger than
$\sim$3.4nm, followed by a still considerable  \emph{total} order parameter ($\sim$0.4-0.5 of maximum
value at 600$^{\circ}$C). Only for
particles smaller than $\sim$3.4nm, is there a strong
reduction in order.

Figure \ref{FIG:PP_T}(b) shows that the Pt-segregation into the first
external layer is compensated by Pt-depletion in the second layer so
that the composition of the core remains close to ideal. This
is a direct consequence of the values of the surface potential ``iso2''
for the two external layers, similar in magnitude but with opposite sign
(Fig. \ref{FIG:Mu_i} and Tab. \ref{TAB:CEs}). Consequently, as the
core remains close to stoichiometry, its ordered state is affected by surface thermodynamics
less than that previously concluded \cite{Asta05-06,Albe05-07}.

Our model considers only configurational entropy and
neglects other effects. For instance, the effects of
vibrational \cite{avdw:vibrev} or magnetic \cite{RubAbr} entropies
are not included. However, the good agreement between the measured and
calculated bulk transition temperatures (Fig. \ref{FIG:PP_T})
suggests that these corrections are small in the important range $T<600^{\circ}$C.
Furthermore, we have also considered nanoparticles embedded in vacuum rather than in the
polymeric medium typically used. Nevertheless, our surface potential
is similar to previous estimates based on experimental data \cite{Albe05-07}
and, more importantly, our most crucial observation pertains to
the second subsurface atomic layer, which is clearly less sensitive
to the surrounding media.
Therefore, our findings reopen the possibility that the weak ordering observed
experimentally is not due to fundamental thermodynamic limitations
but probably to kinetic effects \cite{Exper-KIN} that may be
easier to control \cite{RCWB-05}.

This research was supported by
ONR (N00014-07-1-0878, N00014-07-1-1085, N00014-09-1-0921),
NSF (DMR-0639822), and
DOE (DE-FG07-071D14893).
Computational support was provided by the TeraGrid resources at
TACC (MCA-07S005), NCSA and SDSC (TG-DMR050013N).


\begin{thebibliography}{*}
\bibitem{Exper-KIN} A possible experimental confirmation of the
conclusion can be made based on the data of \cite{Liu-Harrell}, in which it
was shown that the atomic order can be increased by NaCl
treatment, ion irradiation, and by increase of temperature and/or
annealing time, indicating underlying kinetic issues.
\end{thebibliography}
\end{document}